\documentclass[sn-standardnature]{sn-jnl}% Standard Nature Portfolio Reference Style
\jyear{2021}%

\theoremstyle{thmstyleone}%
\theoremstyle{thmstyletwo}%

\theoremstyle{thmstylethree}%

\newcommand{\DNN}{DNN}

\newcommand{\BNN}{MAUL}

\newcommand{\DFT}{DFT}

\newcommand{\GP}{GP}
\newcommand{\np}{neural potential}
\newcommand{\NP}{NP}
\newcommand{\AEV}{AEV}

\begin{document}

\title[Providing Machine Learning Potentials with High Quality Uncertainty Estimates]{Providing Machine Learning Potentials with High Quality Uncertainty Estimates}


% We can update this as needed! Just don't want to forget anyone
\author*[1]{\fnm{Zeynep} \sur{Sumer}}\email{zsumer@ibm.com}
\author[1]{\fnm{James L.} \sur{McDonagh}}
\author*[1]{\fnm{Clyde} \sur{Fare}}\email{clyde.fare@ibm.com}
\author[2]{\fnm{Ravikanth} \sur{Tadikonda}}
\author[2]{\fnm{Viktor} \sur{Z\'olyomi}}
\author[2]{\fnm{David} \sur{Bray}}
\author[1]{\fnm{Edward} \sur{Pyzer-Knapp}}

\affil[1]{\orgdiv{IBM Research Europe - UK}, \orgaddress{\street{Hartree Centre, SciTech Daresbury}, \city{Warrington}, \state{Cheshire}, \postcode{WA4 4AD}, \country{UK}}}
\affil[2]{\orgdiv{The Hartree Centre}, \orgaddress{\street{STFC Daresbury Laboratory}, \city{Warrington}, \state{Cheshire}, \postcode{WA4 4AD}, \country{UK}}}


\abstract{Computational chemistry has come a long way over the course of several decades, enabling subatomic level calculations particularly with the development of Density Functional Theory (DFT). Recently, machine-learned potentials (MLP) have provided a way to overcome the prevalent time and length scale constraints in such calculations. Unfortunately, these models utilise complex and high dimensional representations, making it challenging for users to intuit performance from chemical structure, which has motivated the development of methods for uncertainty quantification. One of the most common methods is to introduce an ensemble of models and employ %utilising 
an averaging approach to determine the uncertainty. In this work, we introduced Bayesian Neural Networks (BNNs) for uncertainty aware energy evaluation as a more principled and resource efficient method to achieve this goal. The richness of our uncertainty quantification enables a new type of hybrid workflow where calculations can be offloaded to a MLP in a principled manner.}


\maketitle
\clearpage
\section{Introduction}\label{sec:intro}
In the past half century, chemical sciences have been augmented by the addition of predictive computational tooling; capable of insight at the most fundamental chemical level of electronic interactions.\cite{kocer2022neural, frenkel2023understanding, sholl2022density} Methods in computational chemistry including \textit{ab-initio} wave function methods and Density Functional Theory (\DFT{}) have enabled molecular and material design applications to become expansive tools used industrially and academically. Recent improvements on computational technologies %resources
have helped accelerated the design of chemical compounds to the point that they can be %are 
specifically tuned for each application use case.\cite{leszczynski2022practical} For example, force fields or semi-empirical quantum chemical methods enabled the study of chemical dynamics and this has been very influential in areas such as drug discovery.\cite{drug_design,semi_empirical} 
Nevertheless, these methods are still too computationally costly to deploy on large-scale screening of chemical space necessary %that are required
for discovery of new materials.\cite{unke_machine_2021}
Hence, new methods are required that reduce the computational expense  while maintaining or improving the accuracy of result. 
%Nevertheless, these methods could not overcome the computational cost of large screenings that are required for discovery of new materials.\cite{unke_machine_2021}
%Overall, suitable compromise methods allowing both speed and accuracy have remained elusive during this time.

%In the past few years, examples of such methods have been described through the combination of fundamental quantum chemical calculation data sets and deep learning. 
One promising technique is through the combination of fundamental quantum chemical calculation data sets and deep learning. 
These methods are now commonly referred to as Machine Learning Interatomic Potentials (MLIP).\cite{anstine2023machine} A subgroup of these potentials, namely Neural Network Potentials (NP), %NNP or NP),
are trained machine learning models which are capable of ingesting chemical information, in the form of a molecular configuration or sum of atomic contributions, and accurately estimating the potential energy of the molecule in seconds as opposed to the hours taken using conventional quantum chemistry techniques.\cite{behler_generalized_2007}
The \NP{}s have been in the literature for more than two decades and come in a variety of forms, including Gaussian Processes (\GP{})\cite{mcdonagh2019utilizing, bartok2010gaussian,deringer2021gaussian,erhard2022machine} and Deep Neural Networks (\DNN{})\cite{smith2017ani,behler2021four,schutt2018schnet}. There are two main types of \NP{}s, namely descriptor-based and end-to-end \NP{}s, which evaluate the molecular information differently. 
A detailed history of their evolution can be found elsewhere.\cite{behler2021four,unke_machine_2021}

%The \NP{}s have been in the literature for more than two decades. The two main types of \NP{}s, namely descriptor-based and end-to-end \NP{}s, evaluate the molecular information differently, 
%a detailed history of their evolution can be found elsewhere.\cite{behler2021four,unke_machine_2021}

\begin{figure}
\centering
\includegraphics[width=12cm]{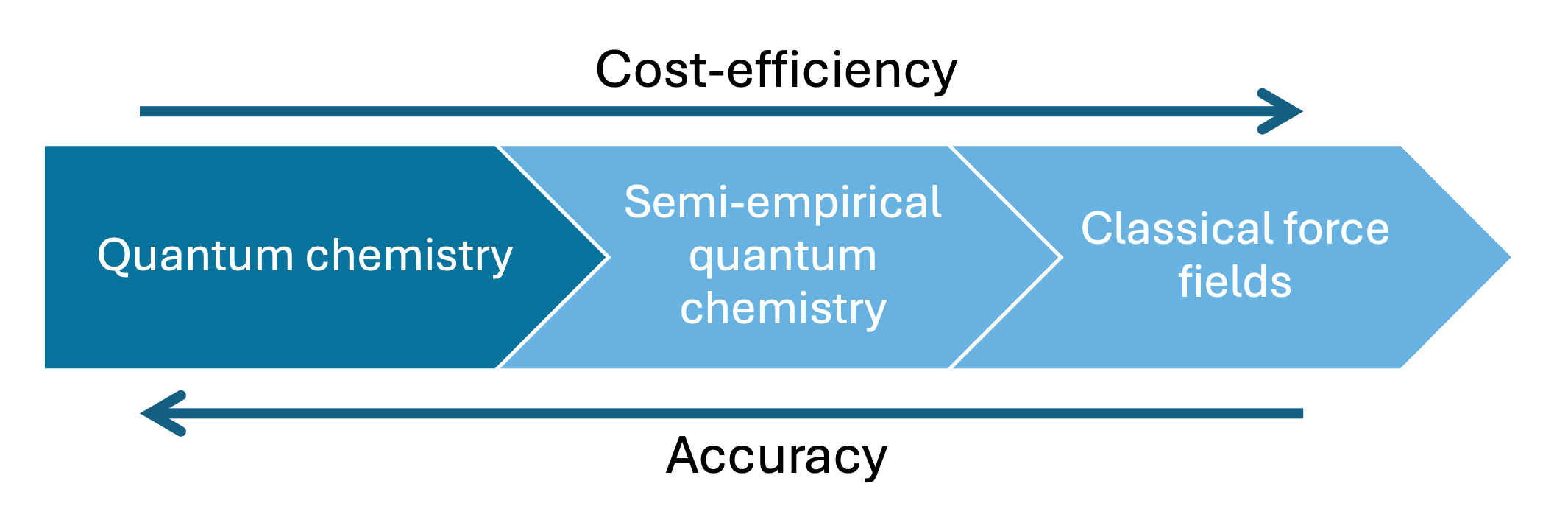}
\caption{The cost/accuracy trade off in simulation approaches.}\label{fig:}
\end{figure}

%These \NP{}s, come in a variety of forms including Gaussian Processes (\GP{})\cite{mcdonagh2019utilizing, bartok2010gaussian,deringer2021gaussian,erhard2022machine} and Deep Neural Networks (\DNN{})\cite{smith2017ani,behler2021four,schutt2018schnet}, typically tuned for specific chemical applications.

The ANI potentials family (short for \textit{accurate neural network engine for molecular energies}) is an encoder-based \NP{} developed by Isayev et. al.\cite{smith2017ani} In ANI potentials each element is described by a dedicated neural network where atomic environment vectors (\AEV{}s) determine the potential energy of an atom in given surroundings and the sum of these atomic energies gives the total energy of the molecule. The authors of ANI acknowledge the need to quantify uncertainty, and have achieved this themselves via ensembling. The uncertainty from these ensembles has found utility in identifying areas where the ANI \np{} may need further training.\cite{smith2018less} 

In order to make learning more efficient, NPs often exploit the invariances inherent in physical systems.\cite{unke_machine_2021} Multiple approaches to uncertainty quantification for neural networks exist, including dropout,\cite{wen2020uncertainty} single Bayesian layers,\cite{fiedler2023improved} ensembles,\cite{smith2020ani} anchored ensembles,\cite{pearce2018uncertainty} GP Layers,\cite{de2024thin} neural GPs,\cite{lee2017deep} and full Bayesian networks.\cite{snoek2012practical}

\begin{table}
\begin{center}
\caption{Uncertainty quantification methods, advantages and disadvantages.}\label{tab2}
\begin{tabular}{ |c|c|c| } 
 \hline
\ Method & Advantage & Disadvantage\\
\hline
\ Ensemble & robustness & non-informative uncertainty\\ 
\ Anchoring & ameliorated uncertainty & reduced accuracy\\ 
\ Dropout & prevented over fitting & computational cost\\ 
\ Bayesian layer & last layer uncertainty & over fitting of latent layer\\ 
\hline

\end{tabular}
\end{center}
\end{table}

In the present work, our contribution is to expand this set of methods to Bayesian Neural Networks (BNN).  We called our new model variant Modified ANI with Uncertainty Limits (MAUL) following the naming theme of these methods. Utilizing BNNs provides the opportunity to not only train models with excellent mean predictive accuracy but to simultaneously provide well founded uncertainty estimates. Here a BNN allows for a single model to provide both an excellent mean predictive performance together with a well founded uncertainty estimate. We demonstrate that such an uncertainty estimate can be used to successfully bound the overall error from a set of predictions.

\section{Results and Discussion}\label{sec3}
\subsection{Energy Calculations}
\begin{figure}
\centering
\includegraphics[width=\textwidth]{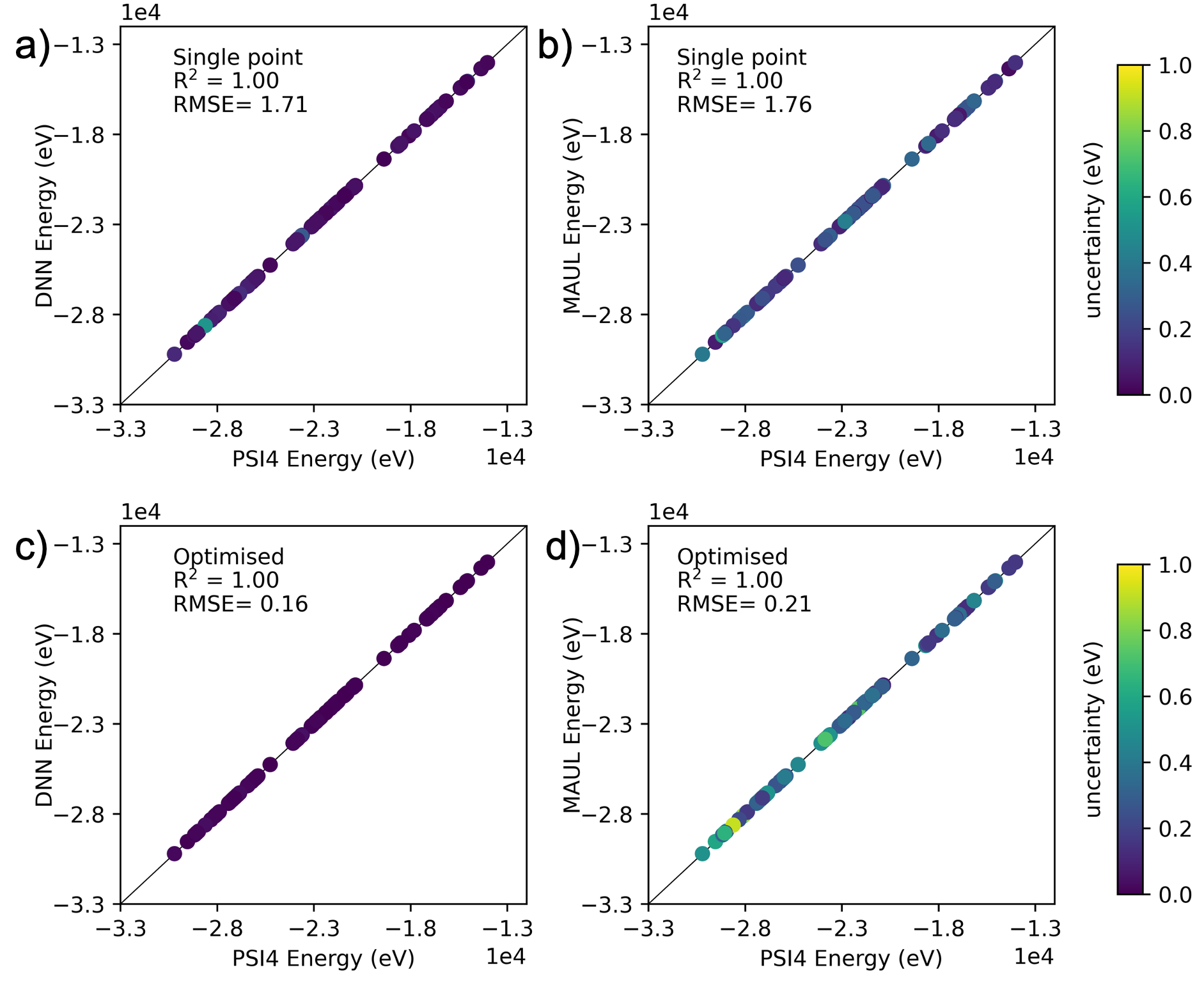}
\caption{
\textbf{a)} \DNN{} ensembles' and \textbf{b)} \BNN{} predictions of single point energies, and \textbf{c)} \DNN{} ensembles' and \textbf{d)} \BNN{} predictions of optimised geomtery energies, as a function of PSI4 calculations for the total of 318 new configurations.
Black diagonal line shows where $x=y$ to guide the eye.}\label{fig:dnn_vs_psi4}
\end{figure}

We tested both DNN and \BNN{} by screening a small set of molecules taken from PubChem; in Section \ref{sec:Data} more information is provided about the selection. Results shown in  Figure \ref{fig:dnn_vs_psi4} are the mean and the standard deviation of the predictions by (a) 9 different \DNN{} ensembles, and (b) \BNN{} results after 20 Monte Carlo samples for single point energies. The coefficient of determination, $R^2$, was 1 in both cases, RMSE values were 1.71 and 1.76\,eV, respectively. The RMSE values reflect the difference between the PSI4 optimised calculations, therefore it is expected that for single point energy calculations it is greater compared to optimised geometry results. We also compared the energies of optimised geometries in Figures \ref{fig:dnn_vs_psi4} (c) and (d), calculated by the models with that of PSI4. Results are well correlated for both \DNN{} and \BNN{} models, both scored $R^2$ of 1 and their RMSE values were 0.16 and 0.21, respectively. Both models achieved similar performances and made accurate predictions, particularly with optimisation. 

\begin{figure}
\centering
\includegraphics[width=\textwidth]{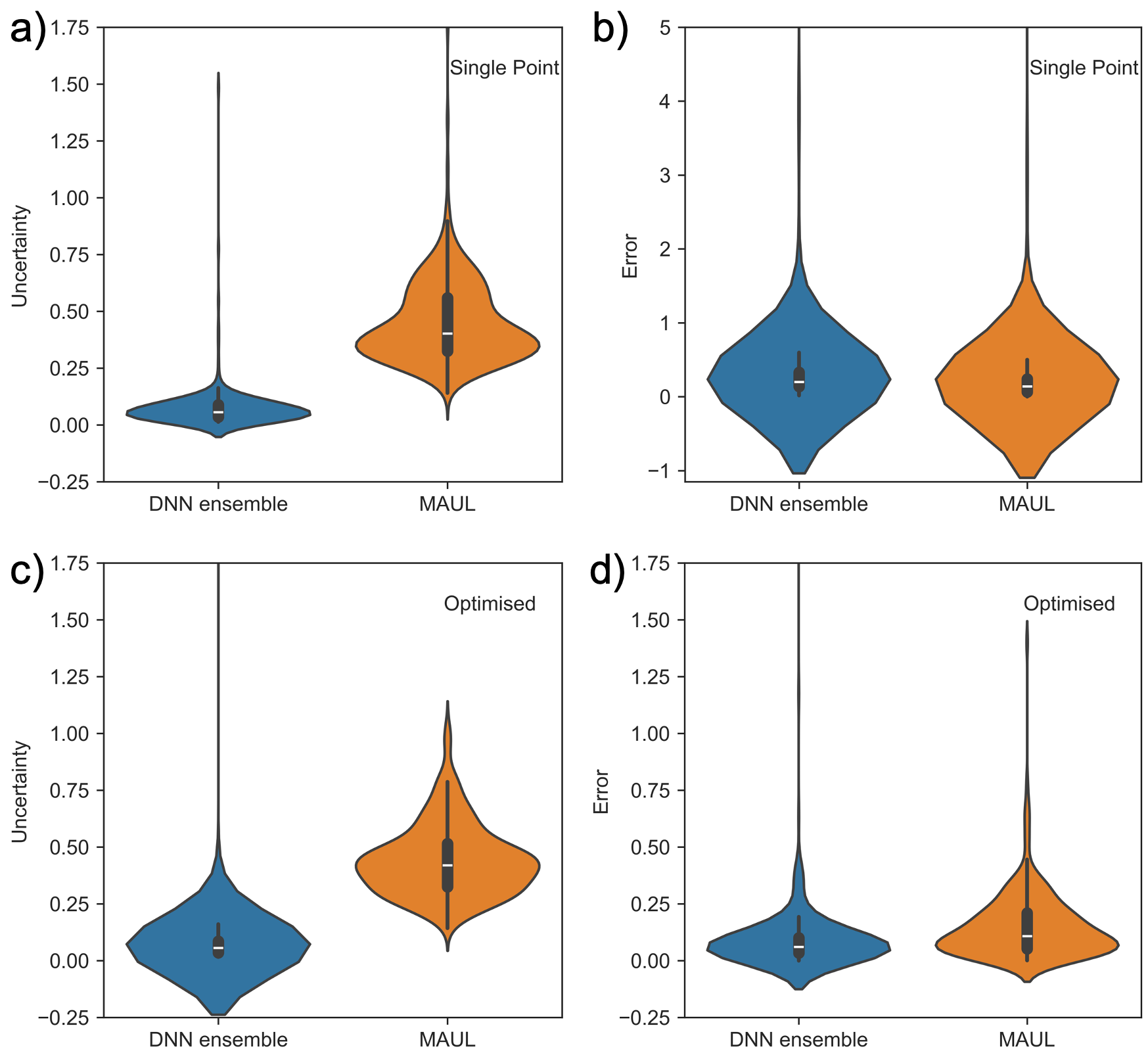}
\caption{Single point geometries presenting \textbf{a)} uncertainty and \textbf{b)} error analysis of DNN ensemble and \BNN{} predictions. Optimised geometries presenting \textbf{c)} uncertainty and \textbf{d)} error analysis of DNN ensemble and \BNN{} predictions.}\label{fig:single_point_std}
\end{figure}

Figures \ref{fig:single_point_std} (a) and (c) depict the standard deviation distribution of each compound in Figure \ref{fig:dnn_vs_psi4} for single point energy and optimised geometry energy, respectively. Although occasionally some DNN ensembles perform poorly, the majority of ensembles produce similar results. Hence, when the ensemble averages are taken the results possess lower standard deviation, on the order of 10$^{-2}$\,eV. As in the example of 2-Methoxycarbonyloxybenzoic acid and toluene, \DNN{} ensemble model would still yield the average single point energy calculated by all ensembles ($-27080.69\pm7.19$\,eV), which is reasonable compared to \BNN{} results ($-27079.21\pm0.65$\,eV) and PSI4 results (-27080.61\,eV), yet the uncertainty would yield the differences in ensembles and not the potential energy of the compounds. 

Figures \ref{fig:single_point_std} (b) and (d) show the difference between the PSI4 calculations and the model predictions for single point energy and optimised geometry energy, respectively. Results show that \DNN{} prediction error is higher than the uncertainty in almost all the cases, whereas the other way around is true for \BNN{}: uncertainties are higher than the errors, covering a secure range for the predictions.

For both \DNN{} and \BNN{} models, errors in single point energy calculations are significant. However, the uncertainty estimation for \DNN{} model in single point energy estimation is the lowest among all measurements. This is not reflective of the errors in the estimation. The uncertainty in ensemble model predictions is strictly dependent on the ensembles and how they are split, bringing the robustness of predictions into question. Building a single neural network that provides the uncertainty, instead of ensembles of models, proposes an efficient approach to energy calculations. All energy estimations alongside the standard deviations shown in this section are available in Supplementary Information.

\subsection{Transition States}
Transition state geometries and energies might be overlooked in geometry optimisation although they contain useful information. Quantifying the uncertainty in the calculations of the transition states is useful in understanding reaction mechanisms, isomerism, chemical binding and so forth. For that purpose we analysed some well-known molecules such as paracetamol, salicylic acid and ethanol. We obtained geometries of minima, transition states (maximum potential energy) and the geometries along the path between minima and the transition state using the TopSearch tool, following the protocol by Dicks et al.\cite{Dicks_TopSearch_2024, dicks2024physics} Note that we did not search for the minimum energy path along the two minima, but let the molecule stretch beyond and calculated the energies of these geometries to see if the uncertainty in the neural potential calculations was grasped.

\begin{figure}
\centering
\includegraphics[width=\textwidth]{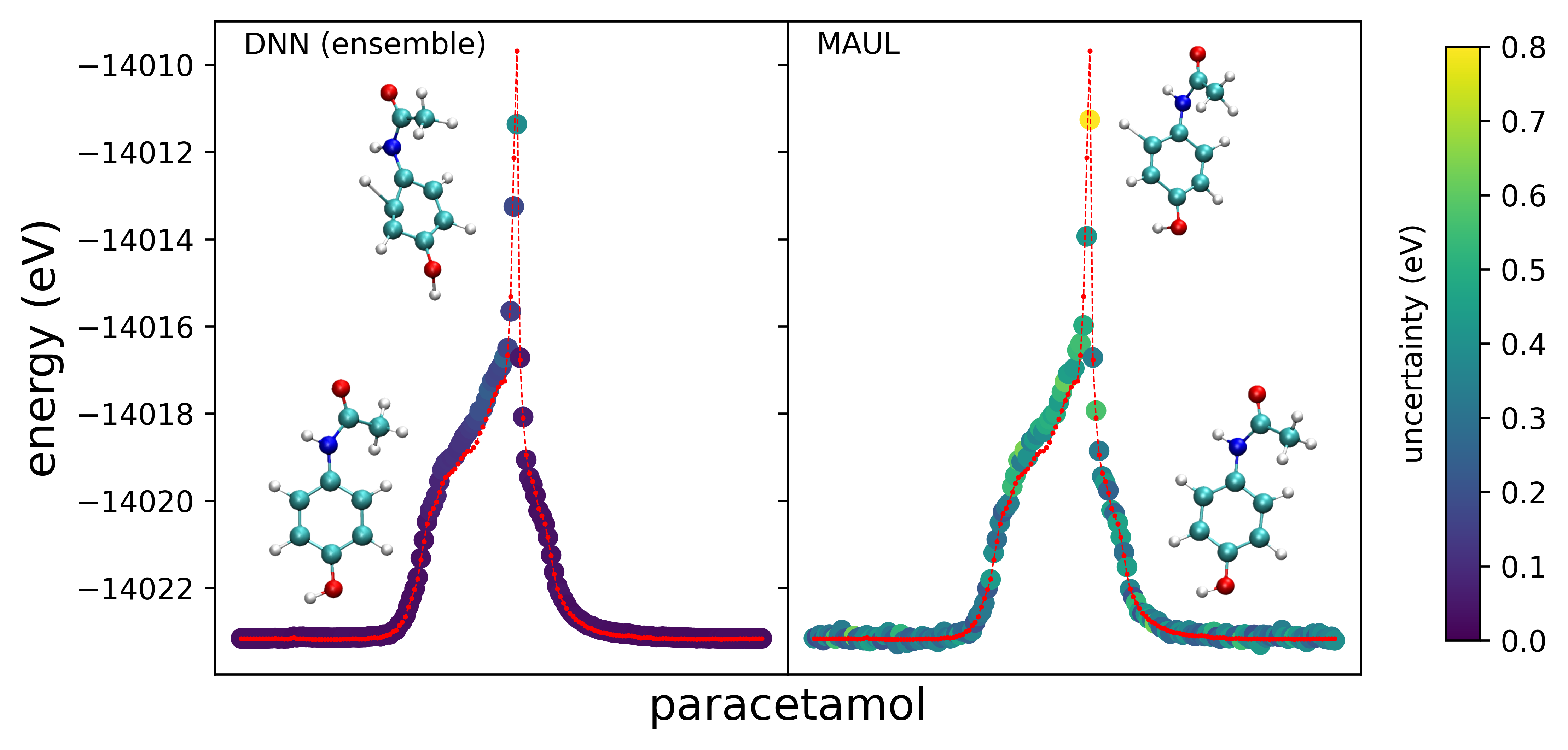}
\caption{An example of the transition of paracetamol molecule between two minima. (Left) \DNN{} ensembles and (Right) \BNN{}. Paracetamol geometries that are shown along the path are valid for both models.Dashed red line represents the PSI4 calculation results.}\label{fig:transition}
\end{figure}

Figure \ref{fig:transition} reveals the energy transition of paracetamol from one minimum to another.\cite{humphrey1996vmd} Both \DNN{} and \BNN{} models calculated similar ground state energies, and transition state energies. As expected, the uncertainties increased in the transition state region, which also represented the area of highest error, as compared to the ground truth. \DNN{} ensemble (\BNN{}) had uncertainty of approximately 0.02\,eV (0.4\,eV) around the minima and 0.2\,eV (0.6\,eV) around the transition state region for paracetamol. \BNN{} was able to more clearly express a lack of confidence in this region, affording a more robust and principled route to direct interventions, such as falling back to DFT.

\subsection{Geometry Optimisation}
In this section we contrast the differences in optimised compound geometries generated by DNN ensembles, MAUL and, for reference, PSI4.
We used the same 318 compounds for which we calculated the single point energies. For quantitative metrics we used the root mean square deviations (RMSD) calculated between geometries of the same compound. 

\begin{figure}
\centering
\includegraphics[width=\textwidth]{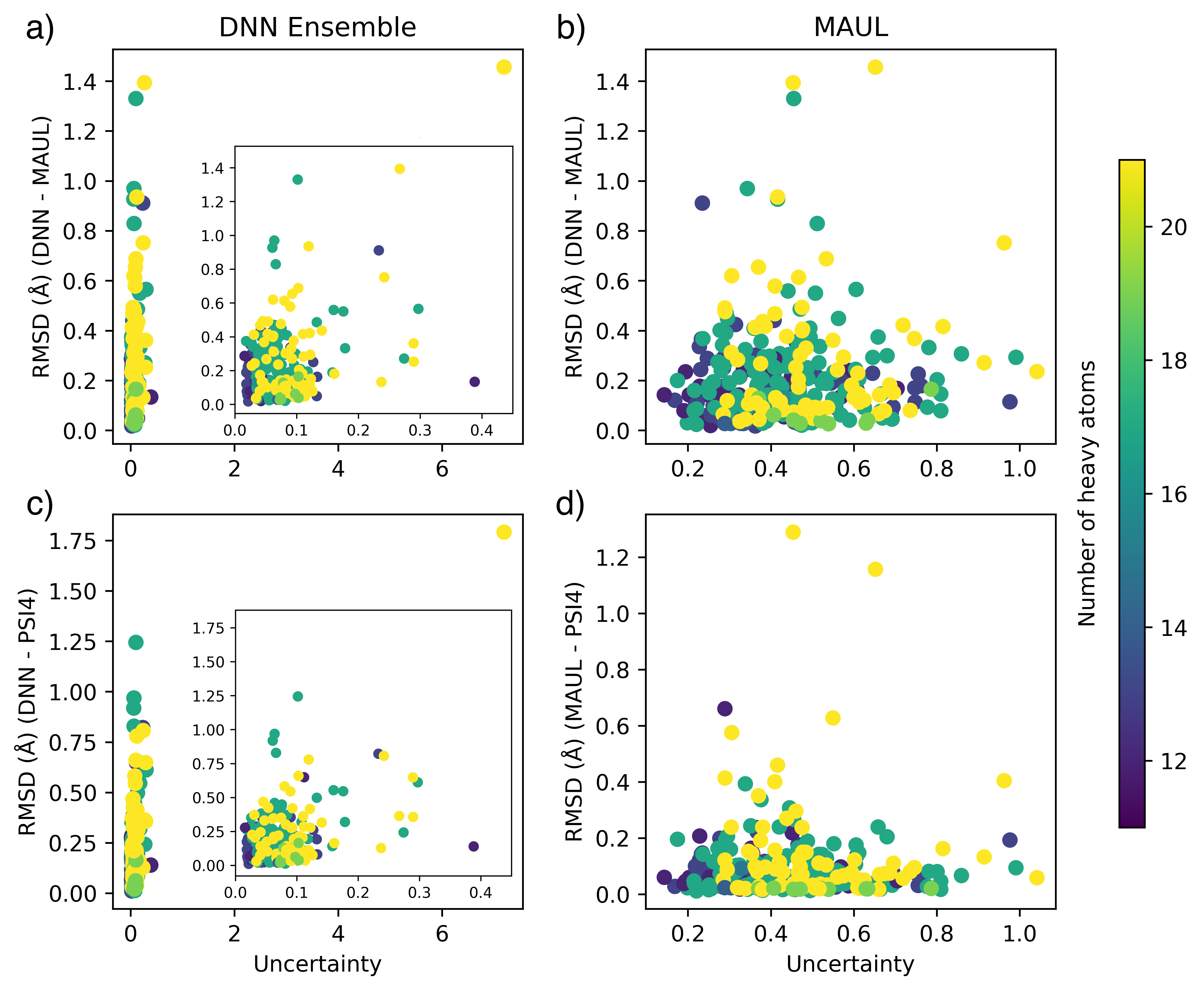}
\caption{RMSD values for the comparison of DNN ensemble and MAUL optimised geometries as a function of \textbf{a)} DNN Ensemble and \textbf{b)} MAUL uncertainties in the predicted energies. RMSD values for \textbf{c)} DNN ensemble - PSI4 geometries and \textbf{d)} MAUL - PSI4 geometries as a function of uncertainties in the energies. Data points are colour-coded with respect to the number of heavy atoms each molecule contain. Insets show zoomed in region of main figure.}\label{fig:geometry}
\end{figure}

The molecules are colour-coded based on the number of heavy atoms they contain. The results reveal that there is a correlation between the uncertainties and the number of heavy atoms in a molecule. As the molecule becomes structurally more complex by addition of each heavy atom, the difference in predictions among different ensembles (or samples) increases and this is reflected on the high uncertainty values. 

Figure \ref{fig:geometry} (a) shows the RMSD values between the DNN ensemble and MAUL optimised geometries as a function of DNN ensemble uncertainty, whereas Figure \ref{fig:geometry} (b) shows the same values as a function of MAUL uncertainty. It is observed that Figure \ref{fig:geometry} (a) possesses a larger uncertainty interval compared to Figure \ref{fig:geometry} (b). This might be the result of the models that are part of the DNN ensemble giving much different results from each other as the molecular complexity increases. As MAUL is always using the same model for different sampling, it is fair to say that the correlation between the high RMSD and MAUL uncertainty is more distinctive due to the richer description of uncertainty. 

In order to assess the prediction accuracy, we compared DNN and MAUL geometries also with the PSI4 optimised geometries, as shown in Figures \ref{fig:geometry} (c) and (d). It is seen that the RMSD values between DNN and PSI4 calculations are similar to the values between MAUL and PSI4 calculations, with some slightly higher values in certain cases. This shows that although in each iteration of the geometry optimisation we will obtain different values with MAUL due to the Bayesian nature of the potential, the optimisation results are still within reliable limits with a correlated uncertainty and therefore provides a reliable geometry optimisation tool. Full set of RMSD results for the set of 318 compounds is available in Supplementary Information.

\subsection{Molecular Dynamics Simulations}

\begin{figure}
\centering
\includegraphics[width=\textwidth]{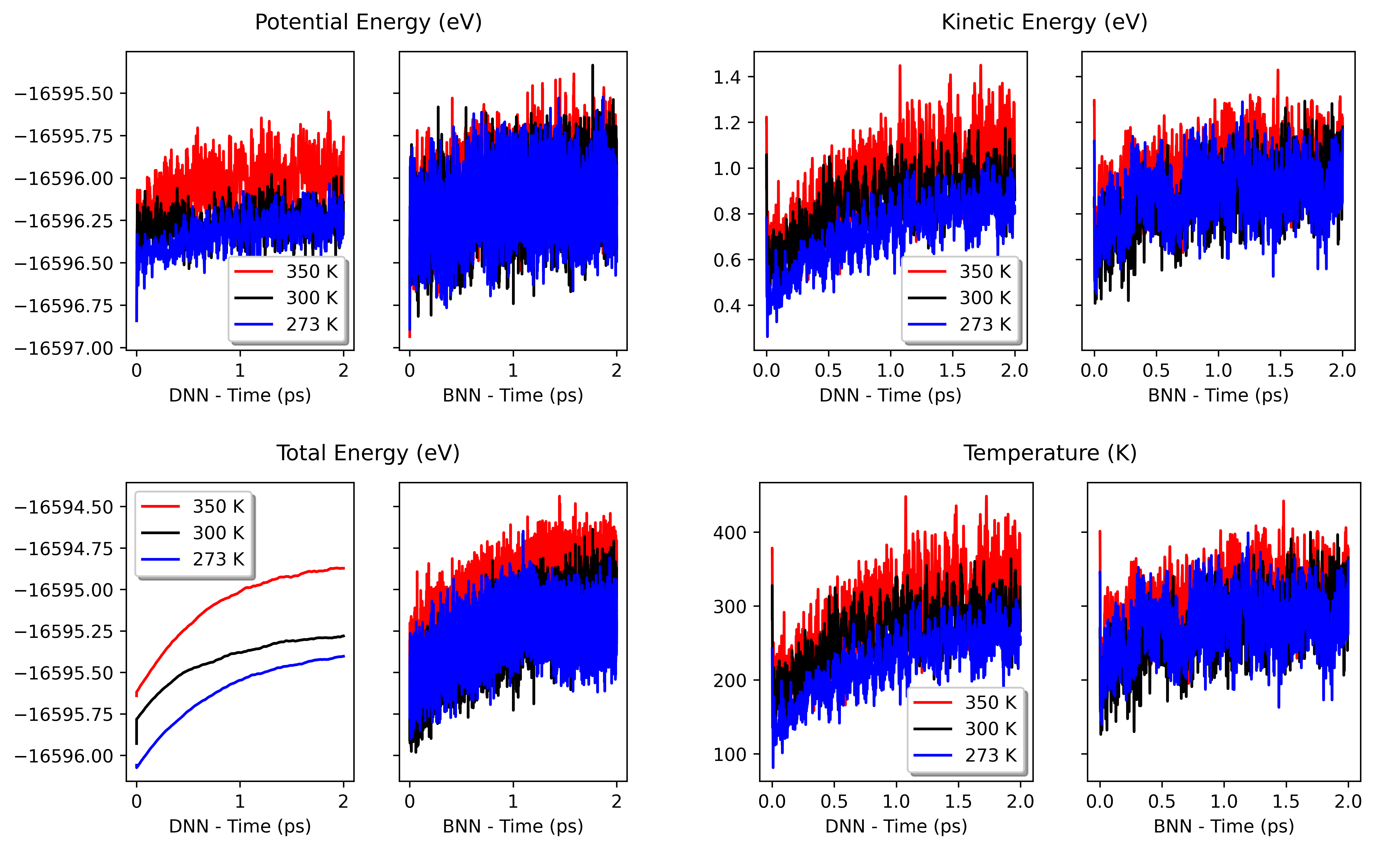}
\caption{Molecular dynamics simulation results for the compound amino (2R)-2-amino-3-phenylpropanoate (CID:69623368). Figure gives four panels % sets Results
showing the potential energy (upper left), kinetic energy (upper right), total energy (lower left), and temperature (lower right) of the system, for simulations with target temperatures of  273\,K, 300\,K and 350\,K. For each panel,  \DNN{} results were shown on left and \BNN{} results were shown on right. Legends are the same for all graphs. Potential energy and total energy graphs include $-1.659$e4, which is equivalent to -16590 eV.}\label{fig:md_all}
\end{figure}

We utilised %took 
the small molecule that was analysed in Section \ref{sec3}.3 (CID: 69623368, amino (2R)-2-amino-3-phenylpropanoate) and performed MD simulations at %using it. We focused on
three different temperatures: 273\,K, 300\,K and 350\,K, %, which are common temperatures for any MD simulation in the literature.
For each temperature, we ran 2\,ps MD simulations under NVT ensemble (see Section \ref{sec:methods}.4 for details). %, which was explained further in Section \ref{sec:methods}.4. 
The results, revealed in Figure \ref{fig:md_all}, suggests that \BNN{} achieves similar accuracy to \DNN{}. Note that all images in Figure \ref{fig:md_all} were treated the same, but the total energy graph for DNN ensemble calculations contain less fluctuation, presumably due to the thermostat and deterministic nature of the potential in calculating the total energy. However the average  results that are obtained by both potentials are similar to each other. Hence,  uncertainty aware neural potentials could provide a viable alternative approach for \textit{ab initio} MD simulations.\cite{wang2023aimd}

\section{Conclusion}\label{sec13}
In this work, we developed an alternative way to provide uncertainty quantification for neural potentials, something that has -to date- only been achieved via ensembling. We built frameworks that allow to calculate the uncertainty in the weights and biases of neural networks, therefore predict molecular energy with an uncertainty in the estimations. We implemented ANI-1x-like potentials, where four elements: C, H, N and O, are represented via symmetry functions and fed through a neural network to calculate many properties such as single point charges and forces.

We calculated the single point energy of 318 new compounds and compared these results to the DFT results for the same compounds. Although mean values were similar, the \BNN{} results provided more robust uncertainty quantification, always giving a reasonable range that covers the DFT results. We also revealed that how the ensembles were separated before the training of the model significantly affects the final result. We compared \DNN{} and \BNN{} performances by \textit{ab initio} MD simulations. These findings show that while DFT is replaced by neural potentials for fast ab-initio MD simulations, uncertainty aware neural potentials can be an alternative to these potentials.

In conclusion, we mimicked well-known ANI-1x potential and created an uncertainty aware neural potential following the same principles. Across a wide variety of calculation types, these uncertainty aware neural potentials in general provided a faster and cheaper alternative. We believe that the results revealed in this paper demonstrate the value of developing models with implicit uncertainty consideration in order to improve their deployment and downstream impact into the computational chemistry and material science communities.

\section{Methods}\label{sec:methods}

\subsection{Network Architecture}
\begin{figure}[H]
\centering
\includegraphics[width=\textwidth]{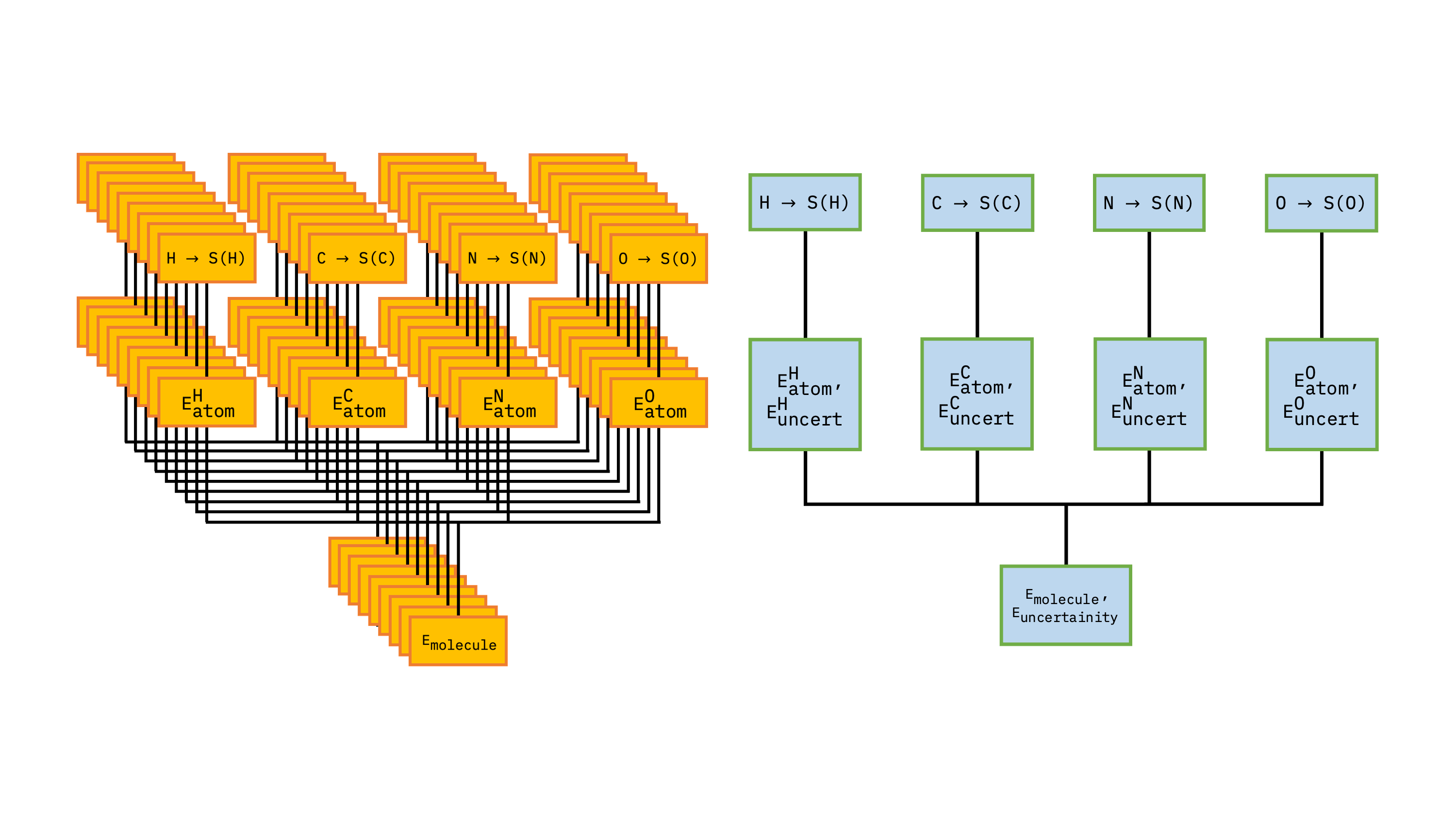}
\caption{Illustration of the concepts of \DNN{} ensembles (left panel) and \BNN{} (right panel). Here $S(X)$ illustrates the application of symmetry functions for molecular featurization.}\label{fig:dnn_bnn_illustration}
\end{figure}

In the present work we compare a popular NP ensemble based on ANI-1x,\cite{smith2018less} to our new model variant named MAUL. Figure \ref{fig:dnn_bnn_illustration} illustrates the difference in the approaches of \DNN{} ensembles and \BNN{}. Traditionally, to analyse statistical errors in neural potentials, chemical configuration data are  split differently
to create multiple training and test subsets.
%differently in order to create different training and test subsets.
For example, the ANI-1 potential consists of 8 different ensembles, each trained separately, and the average of the estimations %was
used to define the error and standard deviation.\cite{smith2017ani} 
Instead, with \BNN{} we propose to obtain estimation errors in a single run which decreases %and, decrease
the computational cost significantly.

Atomic environment vectors, or \AEV{}s, quantitatively represent the local environment of the atoms in a molecule, and they are calculated by using (and modifying) Behler-Parinello symmetry functions.\cite{behler_generalized_2007} We calculated \AEV{}s (${\vec{G}_i^X}$) for each atom \textit{i} following the same protocol used in the ANI-1 potential.\cite{smith2017ani} AEVs are categorised into two chemical environments, radial (${\vec{G}_i^R}$, two-body terms) and angular (${\vec{G}_i^A}$, three-body terms). The local radial environment vector of atom \textit{i} is calculated as follows:

\begin{equation}
%f_C(R_{ij}) = 
%    \begin{cases}
%        0.5\times cos(\frac{\pi R_{ij}}{R_C})+0.5 & \text{for } R_{ij}\leq R_C \\
%        \inline 0.0 & \text{for } R_{ij}>R_C 
%    \end{cases}
G^R_m = 
    \sum_{j \neq i}^N e^{-\eta(R_{ij}-R_s)^2} f_C(R_{ij})
\end{equation}

where \textit{N} represents all atoms in the molecule, \textit{$\eta$} and \textit{$R_s$} are model parameters that change the width and centre of the Gaussian distribution, respectively and \textit{m} is the index number that depends on the model parameters. The pairwise cutoff function \textit{$f_C$} is represented as follows:

\begin{equation}
f_C(R_{ij}) = 
   \begin{cases}
       0.5\times \cos\left(\frac{\pi R_{ij}}{R_C}\right)+0.5 & \text{for } R_{ij}\leq R_C \\
       0.0 & \text{for } R_{ij}>R_C 
   \end{cases}
\end{equation}

where \textit{$R_{ij}$} is the distance between atoms \textit{i} and \textit{j}, and \textit{$R_{C}$} is the cutoff distance.  Size of radial environment vector can be altered by choosing a range of  \textit{$\eta$} and \textit{$R_s$}. Similarly, for the angular environment:

\begin{equation}
\begin{split}
G^{A}_m & = 2^{1-\zeta}\sum_{j,k \neq i}^N (1+\cos\left(\theta_{ijk} - \theta_{s})\right)^\zeta \\
& \times exp\left[-\eta\left(\frac{R_{ij}+R_{ik}}{2}-R_s\right)^2\right]f_C(R_{ij})f_C(R_{ik})
\end{split}
\end{equation}

The angular environment symmetry function also has its own \textit{$\eta$} and \textit{$R_s$} model parameters. In addition to these two, new parameters were included: \textit{$\zeta$} (changes width of peaks), and \textit{$\theta_s$} (arbitrary number of shifts). Detailed analysis of the effect of changing model parameter values can be found elsewhere.\cite{smith2017ani} The default ANI-1x model parameters (and ranges) were used for calculation of 384-dimensional \AEV{}s and the neural networks built with these \AEV{}s for each atom.\cite{smith2018less}

%For the calculation of local environment approximation, cut-off distances were set as 5.2 Å for radial ($R_C_,_R$), and as 3.5 Å for angular ($R_A_,_R$) symmetry functions. We implemented the piece-wise cutoff function as in the reference.\cite{smith2017ani}The Gaussian distribution width parameters ($\eta$, radial and angular) were 16 and 8, respectively. Shift parameters ($R_s$) were changed with step size of 0.9, in 16 steps for radial and 4 steps for angular symmetry functions. Additionally, for angular environment $\Theta_s$ was set between 0.19 - 3.34 (rad) with 8 steps and $\zeta$ (for changing width of the peaks in angular environment) was 32. 

The ANI-1x architecture uses a feed forward neural network for each atom type, this acts on \AEV{}s that encode the molecular environment of a particular atom up to some cutoff distance. Molecular energies are calculated via generating \AEV{}s for each atom, computing an atomic energy associated with each \AEV{} using the appropriate atom network. The atomic energies are then summed to obtain the total energy of the molecule. 

The left panel in Figure \ref{fig:dnn_bnn_illustration} illustrates the idea of a fixed number of \DNN{}s trained on different sub samples of a training data set operating as an ensemble. In this case each \DNN{} independently predicts the energy of an atom and the molecular energy is given by the sum of network outputs. The final predicted molecular energy is given as the ensemble average. An ensemble standard deviation can also be calculated and used as an approximate uncertainty from the ensemble. In contrast, in a \BNN{} (right panel), the uncertainties are obtained for each atomic energy from a single network per species; this is discussed in detail below.

\subsubsection{Initialisation}
Initially we trained standard neural network models by using TorchANI package (version 2.2) for comparison with Bayesian neural networks.\cite{gao_torchani_2020} We used ANI-1x self energies for each atom. We applied a 9-fold cross validation based on different types of components. By doing so, we trained a model with predefined training/validation/test sets and did not randomly split the data. We identified groups as sets of molecules that contain the same number of each atomic species. For example, all molecules containing 9\,C, 12\,H and 6\,O atoms were clustered under the C9H12O6 group, regardless of the structure. Each group has a different number of configurations in the ANI-1x dataset; some might have only three, whereas for other groups this number can go as high as 60 thousand. Controlling the training/validation/test sets enabled a more systematic comparison of models. This approach is anticipated to result in a more robust model. We trained 9 different \DNN{}s each having a different set of 311 groups of components in the validation set and the same 315 groups of components in the test set, as schematically shown in Figure \ref{fig:data_split} (c). We finally trained a 10th ensemble for the BNN, by using all components for the training except the last 315 groups for testing.

\subsubsection{Bayesian Neural Network Architecture}
In a Bayesian neural network, the output of the network (given some initial input) changes from a point estimate of a quantity to a probability distribution over that quantity. Blundell et al.\cite{pmlr-v37-blundell15} introduced Bayes by Backprop which achieves this by placing the Gaussian distributions over the weights and the biases of the network, where a single sample from these weight distributions defines a standard neural network. During the forward pass we draw such samples of the weights and compute the target values associated with them as would be done in a standard neural network. By repeating this process (with different samples of the weights) multiple times, we effectively sample the distribution over our target quantity. The key to the learning then becomes updating the means and variances of the weights according to the distribution seen in the training data. This is done by making use of the reparameterisation trick where random values are drawn from unit Gaussians, these initial uninformative values are then transformed by the means and variances of the weights such that they now correspond to samples drawn from the weight distributions themselves. Since we can consider the initial uninformative random samples to be fixed values if we have an appropriate measure for the loss associated with a particular sample we can then use backpropagation to update the values of the means/variances associated with each weight to minimise that loss. 

Blundell et al.\cite{pmlr-v37-blundell15} derived the loss associated with a variational Bayesian approach that minimises the error between the formal application of Bayes rule and the parametric form of a BNN defined above. We make use of a further variation of this introduced by Wen et al.\cite{wen2018flipout} known as flipout that allows for efficient parallelisation of samples. Within Bayes by Backprop we must iteratively draw samples from the weight distributions then compute the forward pass using them. To accurately sample the posterior we must repeat this many times and this must be done in serial. In particular if we simply generate a batch of the same input and pass this batch through the network, we will simply obtain a batch of identical output - for a single sample of the weights will transform the entire batch identically. To improve this Wen et al.\cite{wen2018flipout} switched to making use of the mean weight values alone to compute the pre-exponential outputs then introduce pseudo-independent weight perturbations of these outputs using the variance weights. This allows for parallel batching of identical inputs to be used to sample the posterior associated with those inputs. Here we make use of the flipout algorithm as implemented within Bayesian Torch.\cite{krishnan2022bayesiantorch}.

Figure \ref{fig:dnn_bnn_illustration} shows a summary of the network. The networks' weights (and biases) are described by a Gaussian probability distribution specified by a mean ($\mu$) and a standard deviation ($\sigma$). Reparametrisation of weights (and biases) were performed by computing the forward and backward passes, transforming initial white Gaussian samples into samples of the weights using the $\mu$/$\sigma$ values associated with each weight. The sampled weights were then used to compute the forward pass, with backpropagation used to compute the gradients with respect to $\mu$ and $\sigma$ values.

\subsection{Datasets}\label{sec:Data}
We used the ANI-1x dataset that was generated for the ANI-1x potential.\cite{smith2020ani} The ANI-1x dataset is composed of approximately 5\,million configurations obtained by utilising active learning sampling techniques and quantum chemical calculations.\cite{smith2018less} These configurations belong to 3114 different groups that are composed of any of the four elements C, H, N and O, with different numbers and combinations of these elements. Among various properties that were saved within the database we utilised the atomic numbers, atomic positions (\textit{coordinates}) and total energy (\textit{wb97x\_dz.energy}) parameters.

\begin{figure}
\centering
\includegraphics[width=\textwidth]{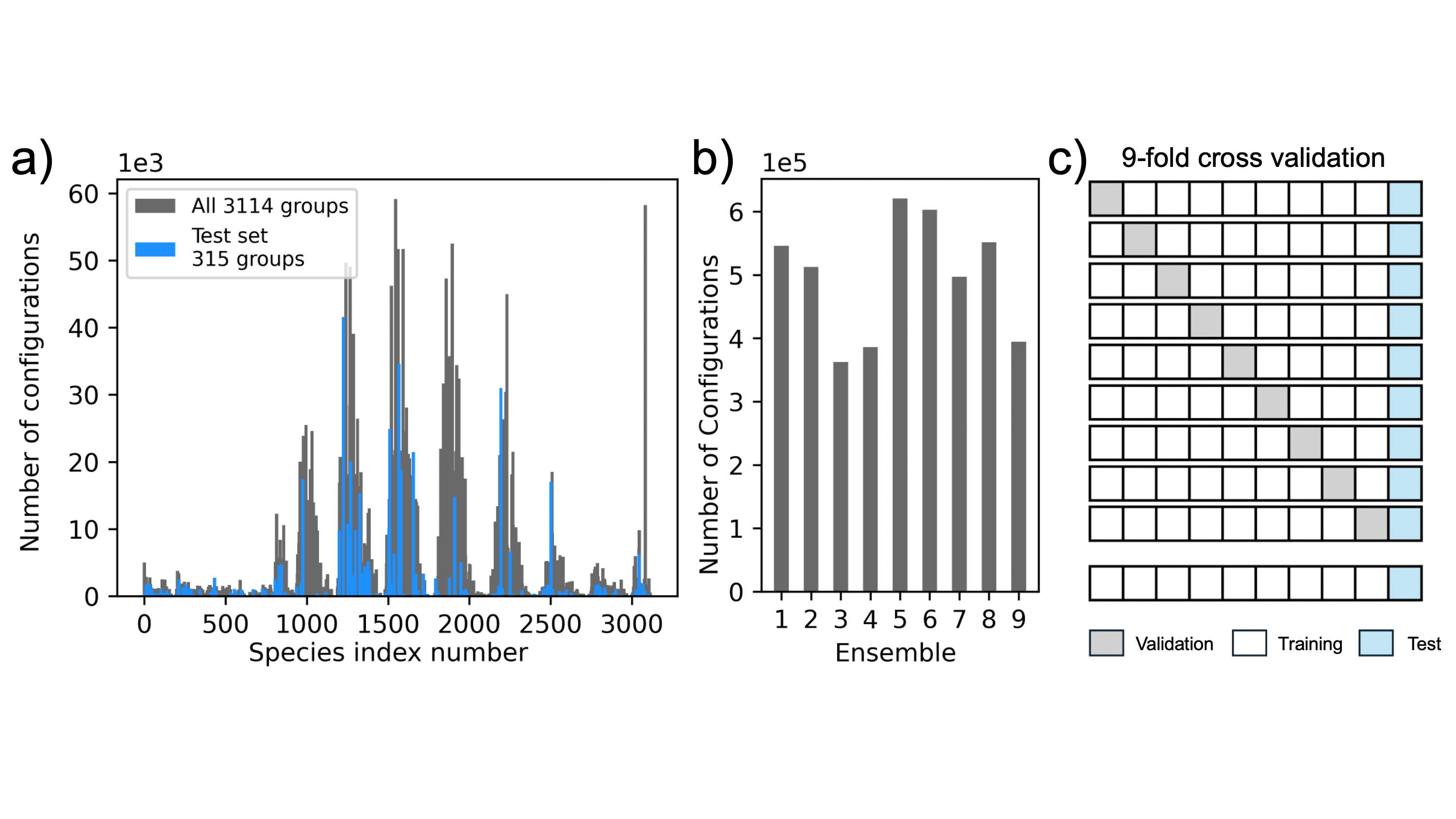}
\caption{\textbf{a} Histogram of number of configurations per groups. Total number of configurations is 4,956,005 (gray bars). Blue bars represent the 315 groups selected for test set (total number of configurations is 481,772). \textbf{b} Total number of configurations per ensemble for validation of the models during training. Each bar represents 311 groups. \textbf{c} Schematic representation of 9-fold cross validation process.}\label{fig:data_split}
\end{figure}

The training, validation and test ensembles for model generation were manually chosen. Figure \ref{fig:data_split} (a) depicts a histogram where the number of configurations for each group can be observed. Separating the sets by configurations that belong to same group can create a bias for groups that have many configurations available, as these data can be distributed across all sets. Therefore, the split is made based on number configurations per groups. 315 groups were chosen for testing of the model. Remaining data was split into 9 equal sets with 311 groups in each, and the training of the model was repeated 9 times by 8 of these subsets, each time using a different subset for testing. Number of configurations per subset (test ensemble) is shown in Figure \ref{fig:data_split} (b).

For testing MAUL against DFT reference data on a small set of molecules taken from PubChem, i.e. a small set of molecules that were not part of the ANI-1x data set, we used the PSI4 code to compute single point energies by DFT. We used the $\omega$B97x functional with a 6-31G* basis set, 99 radial and 590 spherical points on the real space integration grid, robust pruning scheme, and $10^{-6}$ convergence criteria set for both the energy and the density, the same parameters that were used for the generation of the ANI-1x data set. Geometry optimization was done with the BFGS algorithm implemented in the Atomistic Simulation Environment (ASE), which was used also for optimization of the geometries using the neural potentials.\cite{larsen2017atomic}

%\subsection{Metrics}

\subsection{Training}
To train the DNN we made use of the same hyperparameters specified within the original ANI-1 paper.\cite{smith2017ani} Initial learning rate for the models was set to $10^{-6}$, with a learning scheduler set for early stopping rate of $10^{-9}$. The AdamW optimiser in PyTorch (1.12.1) was used with parameter scheduler (\textit{lr\_scheduler}) in order to adjust the learning rate. Validation loss was calculated by MSE between true energies (DFT results) and predicted energies, with the reduction set to none. We used 9-fold cross validation as observed in Figure \ref{fig:data_split} (c). For the discussion of results we only focused on the test score, which was the unincorporated set of compounds (315 groups) discussed in Section \ref{sec:Data}. 

We used a trained DNN model with no validation set (bottom row in Figure \ref{fig:data_split} (c)) as an input for BNN training: we initialised the mean weights of BNNs using the trained DNN weights, and set the initial variance weights to 20\% of the value of the mean. We then adopted a simplified training scheme. When computing the loss associated with a given input, we drew single samples for the output and computed the loss using the simple MSE loss (not including the normal KL divergence term). When updating the weights we first froze the mean weights, updating only the variance weights until we saw convergence of the training error. Following this we then froze the variance weights and updated the means weights until we again saw convergence of the training error. We iteratively repeated training of variances and weights, stepping down the training lengthscale after a round of training both yielded no further progress. We made use of early stopping using validation MSE and the uncertain calibration metrics within the uncertainty toolbox.\cite{chung2021uncertainty} Note the lack of the KL term means this diverges from the formal variational Bayesian treatment as the influence of the prior is only through the initial weights, though our training scheme means this influence remains significant on the final posterior.

Results in the main text only focus on additional molecules that were taken from PubChem, whereas training results and data can be found in Supplementary Information.

\subsection{Molecular Dynamics}
Molecular dynamics simulations were run with Atomistic Simulation Environment (ASE), and MD modules within.\cite{larsen2017atomic} We used Maxwell-Boltzmann distribution in order to set the atomic momenta initially. We run the simulations in NVT ensemble by applying Berendsen thermostat. Timestep was set to 1\,fs for all simulations, time constant for thermostat coupling was chosen as 1\,ps and the simulations were run for 2\,ps. The simulations were run on a single core with a A100-SXM4-40GB GPU.

\section{Data Availability}\label{sec:dataavailability}
ANI-1x Data Set is used for model training, and is publicly available  at the reference cited.\cite{smith2020ani} All codes to generate Bayesian Neural Networks, the model weights and the results for 318 compounds used for model validation are available at: http://github.com/IBM/mod-ani-ul

\backmatter

\bmhead{Acknowledgments}
This work was supported by the Hartree National Centre for Digital Innovation, a collaboration between STFC and IBM.

\section*{Declarations}

%Some journals require declarations to be submitted in a standardised format. Please check the Instructions for Authors of the journal to which you are submitting to see if you need to complete this section. If yes, your manuscript must contain the following sections under the heading `Declarations':

%\begin{itemize}
%\item Conflict of interest/Competing interests (check %journal-specific guidelines for which heading to use)
%\item Ethics approval 
%\item Consent for publication
%\item Availability of data and materials
%\item Authors' contributions
%\end{itemize}

%\noindent
%If any of the sections are not relevant to your manuscript, please include the heading and write `Not applicable' for that section. 


\bibliography{sn-bibliography}% common bib file


\end{document}